\documentclass[10pt,a4paper]{article}

\usepackage[a4paper,top=20mm,bottom=20mm,left=21mm,right=21mm]{geometry}
\usepackage{mathptmx}

\usepackage{amsmath,amssymb,amsfonts,mathtools}
\usepackage{bm}
\usepackage{slashed}
\usepackage{graphicx}
\usepackage{booktabs}
\usepackage{tabularx}
\usepackage{caption}
\usepackage{subcaption}
\usepackage[compress]{cite}
\usepackage{xcolor}
\usepackage{fancyhdr}
\usepackage[colorlinks=true,linkcolor=blue!70!black,citecolor=blue!70!black,urlcolor=blue!70!black]{hyperref}

\begin{document}

\begin{center}
    {\Large\textbf{Elastic Scattering of Dirac Dark Matter with Charged Leptons through a Scalar Mediator: Analytical Framework, Numerical Stability, and Direct-Detection Bounds}}\\[8pt]
    {\normalsize\textbf{Nguyen Nhu Xuan$^1$, Vu Toan Thang$^2$}}\\[3pt]
    {\footnotesize\textit{$^1$Faculty of Physics and Chemical Engineering, Le Quy Don Technical University,\\
    236 Hoang Quoc Viet Street, Nghia Do Ward, Hanoi, Vietnam\\
    $^2$Faculty of Basic Science, Ngo Quyen University, 229B Bach Dang Street, Thu Dau Mot Ward, Ho Chi Minh City, Vietnam}\\
    \texttt{Xuannn@lqdtu.edu.vn} \quad|\quad ORCID: \href{https://orcid.org/0000-0001-7324-9972}{0000-0001-7324-9972}}
\end{center}

\vspace{2pt}
\begin{abstract}
\noindent
We investigate a leptophilic simplified dark matter model in which a Dirac fermion $\chi$ interacts with Standard Model charged leptons $\ell$ through a real scalar mediator $S$. Starting from the low-energy Yukawa interaction Lagrangian in the fermion mass eigenbasis, we derive closed-form tree-level analytical expressions for the invariant scattering amplitude, the spin-averaged squared matrix element, and the differential and total elastic cross sections. To overcome the numerical loss of significance that arises when evaluating the standard antiderivative difference in the heavy-mediator regime, we rearrange the phase-space integral into positive-definite recoil contributions supplemented by convergent power-series expansions. This analytical formulation smoothly connects the light-mediator and heavy-mediator regimes without relying on Effective Field Theory (EFT) contact approximations, allowing us to establish quantitative validity criteria for the contact limit. Finally, we map the microscopic differential cross section onto the reference electron-scattering cross section and evaluate the resulting coupling constraints using experimental data from SENSEI, DAMIC, SuperCDMS, XENON10, and XENON1T.

\vspace{4pt}
\noindent\textit{Keywords:} Dark matter; simplified models; scalar mediator; leptophilic interactions; electron recoil; direct detection; quantum field theory.

\noindent\textit{PACS Nos.:} 95.35.+d, 12.60.-i, 03.65.-w, 13.85.Dz
\end{abstract}

\vspace{4pt}
\hrule
\vspace{6pt}

\section{Introduction}\label{sec:intro}
Cosmological and astrophysical observations --- including galactic rotation curves, weak and strong gravitational lensing, cosmic microwave background (CMB) anisotropies, and large-scale structure surveys --- consistently indicate that approximately $26.5\%$ of the energy density of the Universe consists of non-baryonic dark matter (DM), whereas ordinary baryons contribute only about $4.9\%$~\cite{Planck2018,PDG2024}. Despite this gravitational evidence, the microscopic particle nature and non-gravitational interactions of dark matter remain open problems in fundamental physics.

Weakly Interacting Massive Particles (WIMPs) with masses in the GeV--TeV range have long been studied as leading candidates because they naturally yield the observed thermal relic abundance through freeze-out~\cite{Bertone,Jungman,Arcadi}. However, stringent exclusion limits from nuclear-recoil direct-detection experiments~\cite{XENONnT2023,LZ2023,PandaX4T2021} and collider searches at the Large Hadron Collider (LHC)~\cite{Boveia} have motivated alternative scenarios. Among them, \emph{leptophilic dark matter}, in which the dark sector couples predominantly to Standard Model (SM) charged leptons ($\ell = e, \mu, \tau$) at tree level rather than to quarks~\cite{Kopp2009,Kopp2014,DEramo2016}, naturally suppresses tree-level nuclear recoils while remaining testable through low-threshold electron-recoil and ionization experiments such as SENSEI~\cite{SENSEI2018,SENSEI2020}, DAMIC~\cite{DAMIC2019}, SuperCDMS~\cite{SuperCDMS2021}, XENON10~\cite{XENON10_2012}, and XENON1T (S2-only)~\cite{XENON1T_S2_2019}.

To describe the kinematics and dynamics of such interactions without committing to a specific ultraviolet (UV) completion, phenomenological studies frequently adopt the framework of Simplified Dark Matter Models~\cite{Alwall2009,Alves2012,Abdallah2015,Abercrombie2020}. Compared with Effective Field Theory (EFT) contact operators, which break down when the momentum transfer $q = \sqrt{-t}$ becomes comparable to or exceeds the mediator mass $m_S$, simplified models retain the dynamical propagator of the mediator and remain applicable across both low- and high-momentum-transfer regimes.

In this paper, we investigate elastic scattering between a Dirac fermion dark matter candidate $\chi$ and charged leptons $\ell$ mediated by a real scalar boson $S$ in the $t$-channel. While automated numerical packages (such as \textsc{CalcHEP} or \textsc{MadGraph5}) can compute tree-level cross sections, explicit analytical formulas clarify the asymptotic scaling behavior and provide an independent benchmark for numerical implementations. Furthermore, when the integrated cross section is evaluated in standard 64-bit IEEE-754 floating-point arithmetic, naive subtraction of the indefinite antiderivative $I(0) - I(t_{\min})$ suffers from severe subtractive cancellation in the heavy-mediator regime $|t_{\min}| \ll m_S^2$. The main objectives and results of the present work are as follows:
\begin{enumerate}\setlength{\itemsep}{1pt}\setlength{\parskip}{0pt}\setlength{\topsep}{2pt}
    \item We derive closed-form tree-level analytical expressions for the invariant scattering amplitude, spin-averaged squared matrix element $\overline{|\mathcal{M}|^2}$, differential cross sections $d\sigma/dt$ and $d\sigma/d\Omega$, and total elastic cross section $\sigma(s)$ for $\chi + \ell \to \chi + \ell$, clarifying the low-energy mass-eigenbasis Lagrangian and its electroweak gauge-invariant UV completion.
    \item We construct a numerically stable positive-term reformulation of the total cross section $\sigma(s)$ and the laboratory-frame K\"all\'en function $\lambda(s, m_\chi^2, m_\ell^2)$, supplemented by convergent power-series expansions for the auxiliary recoil integrals $\Phi_1(x)$ and $\Phi_2(x)$ when $x = |t_{\min}|/m_S^2$ is small. We verify this formulation against 80-digit arbitrary-precision arithmetic across broad kinematic regimes.
    \item We analyze the asymptotic light- and heavy-mediator limits, derive the exact and non-relativistic forms of the EFT suppression ratio $\mathcal{R}_{\mathrm{EFT}} = \sigma_{\mathrm{exact}}/\sigma_{\mathrm{EFT}}$, and establish a dimensionally consistent mapping from $d\sigma_{\chi e}/dt$ to the reference cross section $\bar{\sigma}_e$ to examine experimental bounds from SENSEI, DAMIC, SuperCDMS, XENON10, and XENON1T.
\end{enumerate}

The paper is organized as follows. Section~\ref{sec:framework} defines the simplified model Lagrangian, symmetry assumptions, and Feynman rules. Section~\ref{sec:amplitude} presents the analytical derivation of the amplitude and cross sections together with the numerically stable reformulation. Section~\ref{sec:limits} examines the limiting regimes and the validity conditions of the EFT approximation. Section~\ref{sec:numerical} provides numerical illustrations and multi-regime precision benchmarks. Section~\ref{sec:phenomenology} discusses the connection to direct-detection observables and experimental exclusion limits. Section~\ref{sec:conclusions} summarizes our conclusions.

\section{Theoretical Framework}\label{sec:framework}
We consider a leptophilic simplified model in which the Standard Model is extended by a gauge-singlet Dirac fermion $\chi$ representing the dark matter candidate and a real scalar boson $S$ acting as the $t$-channel mediator. Retaining the dynamical propagator of $S$ allows a unified description of the scattering kinematics from sub-keV atomic momentum transfers up to relativistic regimes.

\subsection{Particle Content, Stability Symmetry, and Gauge Structure}
To ensure that the dark matter candidate $\chi$ is stable on cosmological timescales, we impose an exact discrete $\mathbb{Z}_2$ symmetry under which $\chi$ is odd ($\chi \to -\chi$), while the scalar mediator $S$ and all Standard Model fields are even ($S \to +S$, $\psi_{\mathrm{SM}} \to +\psi_{\mathrm{SM}}$). The quantum numbers of the dark-sector fields are listed in Table~\ref{tab:content}.

\begin{table}[htbp]
    \centering
    \caption{\label{tab:content} Particle content and quantum numbers of the dark-sector fields.}
    \begin{tabular}{ccccc}
        \toprule
        Field & Spin & $\mathrm{SU}(3)_C \times \mathrm{SU}(2)_L \times \mathrm{U}(1)_Y$ & $\mathbb{Z}_2$ Parity & Role \\
        \midrule
        $\chi$ & $1/2$ & $(\mathbf{1}, \mathbf{1}, 0)$ & $-1$ & Dirac fermion dark matter \\
        $S$    & $0$   & $(\mathbf{1}, \mathbf{1}, 0)$ & $+1$ & Real scalar mediator \\
        \bottomrule
    \end{tabular}
\end{table}

Two points regarding the symmetry structure and electroweak gauge invariance should be noted. First, while the $\mathbb{Z}_2$ assignment $\chi \to -\chi$ forbids operators linear in $\chi$ and prevents dark matter decay, the $\mathbb{Z}_2$-even scalar $S$ can in general possess self-interactions ($S^3, S^4$) and Higgs-portal couplings ($S H^\dagger H, S^2 H^\dagger H$) in the scalar potential $V(S,H)$. Because such scalar self-couplings do not enter the tree-level $2 \to 2$ elastic amplitude $\chi + \ell \to \chi + \ell$, we work directly in the physical mass eigenbasis after symmetry breaking with renormalized masses $m_\chi$ and $m_S$. Second, above the electroweak symmetry breaking (EWSB) scale, left-handed lepton doublets $L_{\ell L} \sim (\mathbf{1}, \mathbf{2}, -1/2)$ and right-handed charged-lepton singlets $\ell_R \sim (\mathbf{1}, \mathbf{1}, -1)$ transform differently under $\mathrm{SU}(2)_L \times \mathrm{U}(1)_Y$, so the direct Yukawa operator $S\bar{\ell}\ell$ is not gauge invariant under the unbroken electroweak group. In a gauge-invariant ultraviolet (UV) completion~\cite{Kopp2014,DEramo2016,Abercrombie2020}, the coupling to charged leptons is generated either via scalar-Higgs mixing or from the dimension-five operator
\begin{equation}
    \mathcal{L}_{\mathrm{UV}}^{(5)} = -\frac{c_\ell}{\Lambda_{\mathrm{UV}}}\,S\bigl(\bar{L}_{\ell L} H \ell_R + \mathrm{h.c.}\bigr),
    \label{eq:L_UV}
\end{equation}
which yields the effective coupling $g_\ell = c_\ell v_{\mathrm{EW}} / (\sqrt{2}\Lambda_{\mathrm{UV}})$ once the Higgs doublet acquires its vacuum expectation value $\langle H \rangle = (0, v_{\mathrm{EW}}/\sqrt{2})^T$ ($v_{\mathrm{EW}} \simeq 246\,\mathrm{GeV}$). At momentum transfers $q \ll v_{\mathrm{EW}}$, the scattering dynamics is governed by the low-energy mass-eigenbasis Lagrangian below.

\subsection{Low-Energy Interaction Lagrangian and Feynman Rules}
In the post-EWSB mass eigenbasis, the Lagrangian governing the free propagation and tree-level Yukawa interactions of $\chi$, $S$, and the charged leptons $\ell \in \{e, \mu, \tau\}$ is
\begin{equation}
    \mathcal{L} = \mathcal{L}_{\mathrm{SM}} + \bar{\chi}\bigl(i\slashed{\partial} - m_\chi\bigr)\chi + \frac{1}{2}(\partial_\mu S)(\partial^\mu S) - \frac{1}{2}m_S^2 S^2 + \mathcal{L}_{\mathrm{int}},
    \label{eq:Ltot}
\end{equation}
with the interaction term
\begin{equation}
    \mathcal{L}_{\mathrm{int}} = -g_\chi S\,\bar{\chi}\chi - g_\ell S\,\bar{\ell}\ell.
    \label{eq:Lint}
\end{equation}
Here $g_\chi$ and $g_\ell$ are dimensionless real coupling constants treated as free phenomenological parameters. Throughout this paper, $g_\ell$ denotes the coupling to a general charged lepton $\ell$; when specializing to electron targets ($\ell = e$) in Sections~\ref{sec:numerical} and~\ref{sec:phenomenology}, we denote $g_e \equiv g_{\ell=e}$. Since $[\chi] = [\ell] = 3/2$ and $[S] = 1$ in four spacetime dimensions, the operators in Eq.~\eqref{eq:Lint} have canonical mass dimension four, so the low-energy effective theory is perturbatively renormalizable~\cite{PeskinSchroeder,Schwartz,Srednicki}.

From Eq.~\eqref{eq:Lint}, the momentum-space Feynman rules for the scalar vertices are $S\bar{\chi}\chi \to -ig_\chi \mathbf{1}_{4\times 4}$ and $S\bar{\ell}\ell \to -ig_\ell \mathbf{1}_{4\times 4}$. For space-like momentum transfer ($q^2 = t < 0$), the scalar propagator $\Delta_S(q)$ and the spinor completeness relation for on-shell Dirac spinors $u(p,s)$ obeying $(\slashed{p} - m)u(p,s) = 0$ are
\begin{equation}
    \Delta_S(q) = \frac{i}{q^2 - m_S^2 + i\epsilon}, \qquad \sum_{s=1}^{2} u(p,s)\bar{u}(p,s) = \slashed{p} + m.
    \label{eq:compl}
\end{equation}

\section{Analytical Derivation of Elastic Scattering Cross Sections}\label{sec:amplitude}
\subsection{Relativistic Kinematics and Target-Rest-Frame Invariants}
For the two-body elastic scattering process $\chi(p_1, s_1) + \ell(p_2, s_2) \to \chi(p_3, s_3) + \ell(p_4, s_4)$ with on-shell conditions $p_1^2 = p_3^2 = m_\chi^2$ and $p_2^2 = p_4^2 = m_\ell^2$, we define the Mandelstam invariants
\begin{equation}
    s = (p_1 + p_2)^2, \qquad t = (p_1 - p_3)^2 = (p_2 - p_4)^2, \qquad u = (p_1 - p_4)^2 = (p_2 - p_3)^2,
    \label{eq:mandelstam}
\end{equation}
satisfying $s + t + u = 2m_\chi^2 + 2m_\ell^2$. The two-body phase space is determined by the K\"all\'en function
\begin{equation}
    \lambda(s, m_\chi^2, m_\ell^2) = \bigl[s - (m_\chi + m_\ell)^2\bigr]\bigl[s - (m_\chi - m_\ell)^2\bigr] = s^2 + m_\chi^4 + m_\ell^4 - 2s m_\chi^2 - 2s m_\ell^2 - 2m_\chi^2 m_\ell^2.
    \label{eq:lambda}
\end{equation}
The process is kinematically allowed for $s \ge (m_\chi + m_\ell)^2$.

In the laboratory frame where the target lepton is initially at rest, $p_2 = (m_\ell, \vec{0})$, an incoming dark matter particle with speed $v_{\mathrm{DM}}$ (in units $c=1$) has Lorentz factor $\gamma_{\mathrm{DM}} = (1 - v_{\mathrm{DM}}^2)^{-1/2}$, energy $E_\chi = \gamma_{\mathrm{DM}} m_\chi$, and three-momentum $|\vec{p}_1| = \gamma_{\mathrm{DM}} v_{\mathrm{DM}} m_\chi$. The squared center-of-mass energy is
\begin{equation}
    s = m_\chi^2 + m_\ell^2 + 2\gamma_{\mathrm{DM}} m_\chi m_\ell = (m_\chi + m_\ell)^2 + 2(\gamma_{\mathrm{DM}} - 1)m_\chi m_\ell.
    \label{eq:s_lab}
\end{equation}
Substituting Eq.~\eqref{eq:s_lab} into the factorized form of Eq.~\eqref{eq:lambda} yields the algebraic identity
\begin{equation}
    \lambda(s, m_\chi^2, m_\ell^2) = 4m_\ell^2\bigl(E_\chi^2 - m_\chi^2\bigr) = 4m_\chi^2 m_\ell^2\,\gamma_{\mathrm{DM}}^2 v_{\mathrm{DM}}^2.
    \label{eq:lambda_exact_lab}
\end{equation}
While Eqs.~\eqref{eq:lambda} and~\eqref{eq:lambda_exact_lab} are algebraically identical, they behave very differently in finite-precision arithmetic. When $m_\chi \gg m_\ell$ and $v_{\mathrm{DM}} \ll 1$, the expanded polynomial in Eq.~\eqref{eq:lambda} subtracts terms of order $s^2 \sim m_\chi^4$ to obtain a residual of order $4m_\chi^2 m_\ell^2 v_{\mathrm{DM}}^2$; for $m_\chi \ge 100\,\mathrm{GeV}$ and $v_{\mathrm{DM}} = 10^{-3}$, the ratio $4m_\chi^2 m_\ell^2 v_{\mathrm{DM}}^2 / m_\chi^4 \sim 10^{-19}$ lies below the 64-bit IEEE-754 machine epsilon ($\epsilon_{\mathrm{mach}} = 2^{-53} \simeq 1.11 \times 10^{-16}$), causing the polynomial evaluation to lose all significant digits. Using the factorized product in Eq.~\eqref{eq:lambda_exact_lab} together with $\gamma_{\mathrm{DM}} - 1 = v_{\mathrm{DM}}^2 / [\sqrt{1-v_{\mathrm{DM}}^2}(1 + \sqrt{1-v_{\mathrm{DM}}^2})]$ in Eq.~\eqref{eq:s_lab} avoids this subtractive loss of significance.

\subsection{Tree-Level Scattering Amplitude and Spin Summation}
\begin{figure}[htbp]
    \centering
    \includegraphics[width=0.34\textwidth]{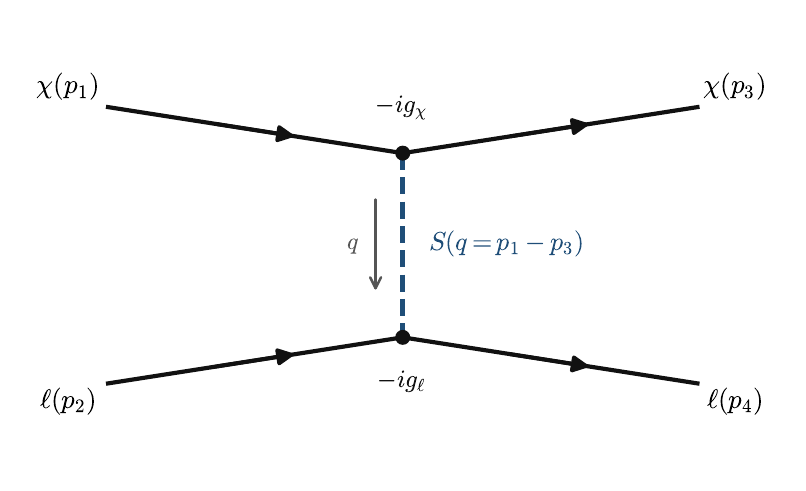}\vspace{-2mm}
    \caption{\label{fig:scattering_diagram} Tree-level Feynman diagram for elastic scattering $\chi(p_1) + \ell(p_2) \to \chi(p_3) + \ell(p_4)$ via $t$-channel exchange of a real scalar mediator $S$ with momentum transfer $q = p_1 - p_3$.}
\end{figure}

The tree-level Feynman diagram for elastic scattering $\chi\ell \to \chi\ell$ via $t$-channel scalar exchange is shown in Fig.~\ref{fig:scattering_diagram}. Applying the Feynman rules of Section~\ref{sec:framework}, the invariant amplitude is
\begin{equation}
    i\mathcal{M} = \bigl[\bar{u}_\chi(p_3, s_3)(-ig_\chi)u_\chi(p_1, s_1)\bigr]\,\frac{i}{t - m_S^2}\,\bigl[\bar{u}_\ell(p_4, s_4)(-ig_\ell)u_\ell(p_2, s_2)\bigr].
    \label{eq:amplitude}
\end{equation}

For unpolarized initial states and unobserved final spins, we average over $s_1, s_2$ and sum over $s_3, s_4$ using Eq.~\eqref{eq:compl}:
\begin{equation}
    \overline{|\mathcal{M}|^2} = \frac{1}{4}\sum_{s_1,s_2,s_3,s_4} |\mathcal{M}|^2 = \frac{g_\chi^2 g_\ell^2}{4(t - m_S^2)^2}\,\mathrm{Tr}\bigl[(\slashed{p}_3 + m_\chi)(\slashed{p}_1 + m_\chi)\bigr]\,\mathrm{Tr}\bigl[(\slashed{p}_4 + m_\ell)(\slashed{p}_2 + m_\ell)\bigr].
    \label{eq:Msq_trace}
\end{equation}
Using the standard trace identities $\mathrm{Tr}(\mathbf{1}) = 4$, $\mathrm{Tr}(\gamma^\mu) = 0$, and $\mathrm{Tr}(\gamma^\mu \gamma^\nu) = 4g^{\mu\nu}$, together with the mass-shell relations $2p_1\cdot p_3 = 2m_\chi^2 - t$ and $2p_2\cdot p_4 = 2m_\ell^2 - t$, we find
\begin{equation}
    \mathrm{Tr}\bigl[(\slashed{p}_3 + m_\chi)(\slashed{p}_1 + m_\chi)\bigr] = 4\bigl(p_1\cdot p_3 + m_\chi^2\bigr) = 2\bigl(4m_\chi^2 - t\bigr),
    \label{eq:trace_chi}
\end{equation}
and $\mathrm{Tr}\bigl[(\slashed{p}_4 + m_\ell)(\slashed{p}_2 + m_\ell)\bigr] = 2(4m_\ell^2 - t)$. Substituting these into Eq.~\eqref{eq:Msq_trace} gives the exact tree-level spin-averaged squared matrix element:
\begin{equation}
    \overline{|\mathcal{M}|^2} = \frac{g_\chi^2 g_\ell^2\,(4m_\chi^2 - t)(4m_\ell^2 - t)}{(t - m_S^2)^2}.
    \label{eq:Msq}
\end{equation}
Since $t \le 0$ in the physical $t$-channel region, both factors $(4m_\chi^2 - t)$ and $(4m_\ell^2 - t)$ are strictly positive.

\subsection{Differential and Total Cross Sections: Numerically Stable Reformulation}
The differential cross section with respect to the Mandelstam variable $t$ is
\begin{equation}
    \frac{d\sigma}{dt} = \frac{\overline{|\mathcal{M}|^2}}{16\pi\,\lambda(s, m_\chi^2, m_\ell^2)} = \frac{g_\chi^2 g_\ell^2\,(4m_\chi^2 - t)(4m_\ell^2 - t)}{16\pi\,\lambda(s, m_\chi^2, m_\ell^2)\,(t - m_S^2)^2}.
    \label{eq:dsdt}
\end{equation}
In the center-of-mass (CM) frame, the three-momentum magnitude is $|\vec{p}_{\mathrm{cm}}| = \sqrt{\lambda(s, m_\chi^2, m_\ell^2)}/(2\sqrt{s})$. Relating $t$ to the CM scattering angle $\theta \in [0, \pi]$ and denoting the maximum momentum transfer by $q_{\max} \equiv 2|\vec{p}_{\mathrm{cm}}|$ (mass dimension $+1$), the physical range $t_{\min} \le t \le 0$ is given by
\begin{equation}
    t = -\frac{\lambda(s, m_\chi^2, m_\ell^2)}{s}\sin^2\!\frac{\theta}{2}, \qquad |t_{\min}| = -t_{\min} = q_{\max}^2 = \frac{\lambda(s, m_\chi^2, m_\ell^2)}{s} = \frac{4m_\chi^2 m_\ell^2\,\gamma_{\mathrm{DM}}^2 v_{\mathrm{DM}}^2}{s}.
    \label{eq:tmin}
\end{equation}
The differential cross section per unit solid angle in the CM frame is
\begin{equation}
    \frac{d\sigma}{d\Omega} = \frac{\lambda(s, m_\chi^2, m_\ell^2)}{4\pi s}\,\frac{d\sigma}{dt} = \frac{g_\chi^2 g_\ell^2\,(4m_\chi^2 - t)(4m_\ell^2 - t)}{64\pi^2 s\,(t - m_S^2)^2}.
    \label{eq:dOmega}
\end{equation}

To integrate Eq.~\eqref{eq:dsdt} over $t \in [t_{\min}, 0]$, we set $A = 4m_\chi^2$, $B = 4m_\ell^2$, and $M^2 = m_S^2$. Using the partial-fraction decomposition
\begin{equation}
    \frac{(A - t)(B - t)}{(t - M^2)^2} = 1 + \frac{2M^2 - A - B}{t - M^2} + \frac{(A - M^2)(B - M^2)}{(t - M^2)^2},
    \label{eq:partial_frac}
\end{equation}
one obtains the indefinite antiderivative
\begin{equation}
    I(t) = t + (2M^2 - A - B)\ln(M^2 - t) - \frac{(A - M^2)(B - M^2)}{t - M^2},
    \label{eq:I_antideriv}
\end{equation}
and the total elastic cross section
\begin{equation}
    \sigma(s) = \frac{g_\chi^2 g_\ell^2}{16\pi\,\lambda(s, m_\chi^2, m_\ell^2)}\bigl[I(0) - I(t_{\min})\bigr].
    \label{eq:sigmatot}
\end{equation}

Although Eq.~\eqref{eq:sigmatot} is exact symbolically, direct evaluation of $I(0) - I(t_{\min})$ via Eq.~\eqref{eq:I_antideriv} in floating-point arithmetic loses precision when the dimensionless ratio $x \equiv |t_{\min}|/m_S^2 = q_{\max}^2 / m_S^2 = \lambda(s, m_\chi^2, m_\ell^2)/(s\,m_S^2)$ is very small ($x \ll 1$). Expanding the three terms of $I(0) - I(t_{\min})$ in powers of $x$ shows that contributions of order $\mathcal{O}(m_S^2 x)$ and $\mathcal{O}((A+B)x)$ cancel analytically between the linear, logarithmic, and rational pieces, leaving a physical cross section proportional to $AB x / m_S^2$. In 64-bit double precision, this cancellation degrades accuracy for $x \lesssim 10^{-6}$ and causes complete loss of significance for $x \lesssim 10^{-12}$.

To avoid inter-term subtraction, we change the integration variable to $u = -t = q^2 \in [0, |t_{\min}|]$ and expand the numerator in positive powers of $u$: $(A + u)(B + u) = AB + (A + B)u + u^2$. Integrating each term over $u \in [0, |t_{\min}|]$ gives the rearranged closed-form expression
\begin{equation}
    \sigma(s) = \frac{g_\chi^2 g_\ell^2}{16\pi\,\lambda(s, m_\chi^2, m_\ell^2)}\left[\frac{16m_\chi^2 m_\ell^2\,|t_{\min}|}{m_S^2\bigl(m_S^2 + |t_{\min}|\bigr)} + 4(m_\chi^2 + m_\ell^2)\,\Phi_1(x) + m_S^2\,\Phi_2(x)\right],
    \label{eq:sigma_stable}
\end{equation}
where the non-negative dimensionless auxiliary recoil integrals $\Phi_1(x)$ and $\Phi_2(x)$ ($x > 0$) and their convergent power-series representations are given by
\begin{equation}
\begin{aligned}
    \Phi_1(x) &\equiv \ln(1 + x) - \frac{x}{1 + x} = \sum_{n=0}^{\infty}\frac{n+1}{n+2}(-1)^n x^{n+2} = \frac{x^2}{1 + x}\sum_{n=0}^{\infty}\frac{(-x)^n}{(n + 1)(n + 2)}, \\
    \Phi_2(x) &\equiv x - 2\ln(1 + x) + \frac{x}{1 + x} = \sum_{n=0}^{\infty}\frac{n+1}{n+3}(-1)^n x^{n+3} = \frac{2x^3}{1 + x}\sum_{n=0}^{\infty}\frac{(-x)^n}{(n + 2)(n + 3)}.
\end{aligned}
\label{eq:Phi_series}
\end{equation}
In Eq.~\eqref{eq:sigma_stable}, all three bracketed terms are positive for $x > 0$, separating the leading scalar contribution from the $\mathcal{O}(q^2)$ and $\mathcal{O}(q^4)$ spinor recoil corrections. Because direct evaluation of $\ln(1+x) - x/(1+x)$ and $x - 2\ln(1+x) + x/(1+x)$ still suffers from internal cancellation of $\mathcal{O}(x)$ and $\mathcal{O}(x^2)$ terms when $x \ll 1$, we use the series forms in Eq.~\eqref{eq:Phi_series} for small arguments.
For numerical implementation in 64-bit floating-point arithmetic, we evaluate $\Phi_1(x)$ and $\Phi_2(x)$ piecewise: for $0 \le x < x_{\mathrm{th}} = 0.5$, we sum the $(1+x)^{-1}$-factored series in Eq.~\eqref{eq:Phi_series} using Horner's rule up to $N_{\mathrm{trunc}} = 50$ terms (where the truncation remainder is bounded by $x_{\mathrm{th}}^{50}/2550 < 10^{-18} \ll \epsilon_{\mathrm{mach}}$), while for $x \ge 0.5$ we evaluate the closed-form expressions in Eq.~\eqref{eq:Phi_series} using the library function $\mathrm{log1p}(x) = \ln(1+x)$.

\section{Limiting Regimes and Validity of the EFT Approximation}\label{sec:limits}
The transition between the contact-interaction regime and the long-range propagator regime is governed by the dimensionless ratio $x = |t_{\min}|/m_S^2 = q_{\max}^2 / m_S^2$, or equivalently by the scaled mediator mass $\eta \equiv m_S / q_{\max} = m_S / \sqrt{|t_{\min}|} = x^{-1/2}$.

\subsection{Heavy-Mediator (EFT Contact) Regime \texorpdfstring{($m_S^2 \gg |t_{\min}|$, $x \ll 1$)}{(m_S^2 >> |t_min|, x << 1)}}
When $m_S^2 \gg |t_{\min}|$ ($\eta \gg 1$), integrating out the heavy scalar $S$ generates the dimension-six four-fermion contact operator $\mathcal{L}_{\mathrm{EFT}} = (g_\chi g_\ell / m_S^2)(\bar{\chi}\chi)(\bar{\ell}\ell)$. Replacing $(t - m_S^2)^2 \to m_S^4$ in Eq.~\eqref{eq:dsdt} and integrating over $t \in [t_{\min}, 0]$ yields the exact tree-level EFT total cross section:
\begin{equation}
    \sigma_{\mathrm{EFT}}(s) = \frac{g_\chi^2 g_\ell^2\,|t_{\min}|}{16\pi\,\lambda(s, m_\chi^2, m_\ell^2)\,m_S^4}\left[16m_\chi^2 m_\ell^2 + 2(m_\chi^2 + m_\ell^2)|t_{\min}| + \frac{1}{3}|t_{\min}|^2\right].
    \label{eq:sigma_EFT}
\end{equation}
Note that keeping the leading term of $\Phi_1(x) \simeq x^2/2$ and $\Phi_2(x) \simeq x^3/3$ in Eq.~\eqref{eq:sigma_stable} reproduces Eq.~\eqref{eq:sigma_EFT} identically. In the non-relativistic limit ($v_{\mathrm{DM}} \ll 1$), where $s \simeq (m_\chi + m_\ell)^2$ and $|t_{\min}| \simeq 4\mu_{\chi\ell}^2 v_{\mathrm{DM}}^2$ with reduced mass $\mu_{\chi\ell} = m_\chi m_\ell/(m_\chi + m_\ell)$, Eq.~\eqref{eq:sigma_EFT} reduces to
\begin{equation}
    \sigma_{\mathrm{EFT}}^{\mathrm{NR}} \simeq \frac{g_\chi^2 g_\ell^2\,\mu_{\chi\ell}^2}{\pi\,m_S^4}.
    \label{eq:sigma_EFT_NR}
\end{equation}

\subsection{Light-Mediator Regime \texorpdfstring{($m_S^2 \ll |t_{\min}|$, $x \gg 1$)}{(m_S^2 << |t_min|, x >> 1)}}
When $m_S^2 \ll |t| \le |t_{\min}|$, the propagator behaves as $(t - m_S^2)^{-2} \simeq |t|^{-2}$, resulting in a forward-peaked differential distribution $d\sigma/dt \propto |t|^{-2}$ regularized at small angles $|t| \lesssim m_S^2$. For $x = |t_{\min}|/m_S^2 \gg 1$, we have $x/(1+x) \simeq 1$, $\Phi_1(x) \simeq \ln x - 1$, and $\Phi_2(x) \simeq x - 2\ln x$, so Eq.~\eqref{eq:sigma_stable} takes the asymptotic form
\begin{equation}
    \sigma_{\mathrm{light}}(s) \simeq \frac{g_\chi^2 g_\ell^2}{\pi s\,m_S^2} + \frac{g_\chi^2 g_\ell^2}{16\pi\,\lambda(s, m_\chi^2, m_\ell^2)}\left[4(m_\chi^2 + m_\ell^2)\ln\!\left(\frac{|t_{\min}|}{m_S^2}\right) + |t_{\min}|\right].
    \label{eq:sigma_light}
\end{equation}
If the kinematics is non-relativistic ($|t_{\min}| \simeq 4\mu_{\chi\ell}^2 v_{\mathrm{DM}}^2 \ll 4m_\ell^2$) and $m_S$ satisfies the hierarchy $m_S^2 \ln(|t_{\min}|/m_S^2) \ll 4\mu_{\chi\ell}^2 |t_{\min}|/(m_\chi^2 + m_\ell^2)$, the first term in Eq.~\eqref{eq:sigma_light} dominates over the logarithmic recoil terms. Holding $v_{\mathrm{DM}}$ fixed in this non-relativistic light-mediator regime yields
\begin{equation}
    \sigma_{\mathrm{light}}^{\mathrm{NR}} \simeq \frac{g_\chi^2 g_\ell^2\,\mu_{\chi\ell}^2}{\pi\,m_S^2\,|t_{\min}|} = \frac{g_\chi^2 g_\ell^2}{4\pi\,m_S^2\,v_{\mathrm{DM}}^2}.
    \label{eq:sigma_light_NR}
\end{equation}
In this regime, the $\mu_{\chi\ell}^2$ factor in the numerator cancels against $|t_{\min}| \simeq 4\mu_{\chi\ell}^2 v_{\mathrm{DM}}^2$ in the denominator, yielding a cross section proportional to $m_S^{-2} v_{\mathrm{DM}}^{-2}$ that becomes independent of $m_\chi$ at fixed $v_{\mathrm{DM}}$ and $m_S$. We note that Eq.~\eqref{eq:sigma_light_NR} describes the asymptotic behavior of the microscopic tree-level cross section; in physical applications, very light scalar mediators ($m_S \lesssim \mathrm{keV}$) are subject to additional astrophysical, cosmological, and in-medium screening constraints.

\subsection{Propagator Suppression Ratio and Condition for EFT Validity}
Taking the ratio of the exact total cross section~\eqref{eq:sigma_stable} to the EFT cross section~\eqref{eq:sigma_EFT} gives the general propagator suppression factor
\begin{equation}
    \mathcal{R}_{\mathrm{EFT}}(s, m_S) \equiv \frac{\sigma_{\mathrm{exact}}(s)}{\sigma_{\mathrm{EFT}}(s)} = \frac{\dfrac{16m_\chi^2 m_\ell^2}{1 + x} + \dfrac{4(m_\chi^2 + m_\ell^2)}{x}\,\Phi_1(x) + \dfrac{m_S^2}{x}\,\Phi_2(x)}{16m_\chi^2 m_\ell^2 + 2(m_\chi^2 + m_\ell^2)|t_{\min}| + \dfrac{1}{3}|t_{\min}|^2}.
    \label{eq:EFT_ratio_exact}
\end{equation}
In the non-relativistic regime where $|t_{\min}| \simeq 4\mu_{\chi\ell}^2 v_{\mathrm{DM}}^2 \ll 4m_\ell^2$, the spinor recoil terms in both the numerator and denominator of Eq.~\eqref{eq:EFT_ratio_exact} are suppressed relative to $16m_\chi^2 m_\ell^2$ by factors of order $|t_{\min}|/(8m_\ell^2) \le v_{\mathrm{DM}}^2 / 2$ (which is $\simeq 5 \times 10^{-7}$ for $v_{\mathrm{DM}} = 10^{-3}c$). Neglecting these $\mathcal{O}(v_{\mathrm{DM}}^2)$ recoil corrections, the suppression ratio and the relative EFT overestimation error $\delta_{\mathrm{EFT}} \equiv (\sigma_{\mathrm{EFT}} - \sigma_{\mathrm{exact}})/\sigma_{\mathrm{exact}}$ simplify to the non-relativistic forms
\begin{equation}
    \mathcal{R}_{\mathrm{EFT}}^{\mathrm{NR}}(\eta) = \frac{1}{1 + |t_{\min}|/m_S^2} = \frac{\eta^2}{1 + \eta^2}, \qquad
    \delta_{\mathrm{EFT}}^{\mathrm{NR}}(\eta) = \frac{|t_{\min}|}{m_S^2} = \frac{1}{\eta^2},
    \label{eq:EFT_ratio}
\end{equation}
with $\eta = m_S / q_{\max} = m_S / \sqrt{|t_{\min}|}$. Within the non-relativistic approximation (accurate to $\mathcal{O}(v_{\mathrm{DM}}^2) \sim 10^{-6}$ at halo velocities), Eq.~\eqref{eq:EFT_ratio} provides three quantitative criteria:
\begin{itemize}\setlength{\itemsep}{1pt}\setlength{\parskip}{0pt}\setlength{\topsep}{2pt}
    \item At the transition scale $\eta = 1$ ($m_S = q_{\max} = \sqrt{|t_{\min}|} \simeq 2\mu_{\chi\ell}v_{\mathrm{DM}}$), $\mathcal{R}_{\mathrm{EFT}}^{\mathrm{NR}}(1) = 0.50$, so the contact approximation overestimates the cross section by $100\%$ ($\delta_{\mathrm{EFT}}^{\mathrm{NR}} = 1$).
    \item Achieving a relative error $\delta_{\mathrm{EFT}}^{\mathrm{NR}} \le 10\%$ requires $\eta \ge \sqrt{10} \simeq 3.16$, or $m_S \ge 3.16\sqrt{|t_{\min}|}$.
    \item Achieving $\delta_{\mathrm{EFT}}^{\mathrm{NR}} \le 1\%$ requires $\eta \ge 10$, i.e., $m_S \ge 10\sqrt{|t_{\min}|} \simeq 20\mu_{\chi\ell}v_{\mathrm{DM}}$.
\end{itemize}

\section{Numerical Illustrations and Precision Benchmarks}\label{sec:numerical}
We now evaluate the analytical expressions of Sections~\ref{sec:amplitude}--\ref{sec:limits} for elastic dark matter--electron scattering ($\ell = e$, $m_e = 0.51099895 \times 10^{-3}\,\mathrm{GeV}$, $g_e \equiv g_{\ell=e}$). The benchmark parameters used in our numerical calculations are listed in Table~\ref{tab:bench}. For $g_\chi = g_e = 0.1$, the loop expansion parameters $g_\chi^2/(4\pi) = g_e^2/(4\pi) \simeq 7.96 \times 10^{-4} \ll 1$ remain well within the perturbative domain.

\begin{table}[htbp]
    \centering
    \caption{\label{tab:bench} Benchmark parameters used in Section~\ref{sec:numerical}.}
    \resizebox{\textwidth}{!}{%
    \begin{tabular}{llll}
        \toprule
        Parameter & Symbol & Benchmark Value / Range & Physical Context \\
        \midrule
        Target electron mass & $m_e$ & $0.51099895 \times 10^{-3}\,\mathrm{GeV}$ & Free or atomic electron target \\
        Halo DM speed & $v_{\mathrm{DM}}$ & $1.0 \times 10^{-3}\,c$ & Standard Halo Model ($v_0 \approx 220\,\mathrm{km/s}$) \\
        Dark-sector coupling & $g_\chi$ & $0.1$ & Perturbative coupling ($g_\chi^2/4\pi \ll 1$) \\
        Electron coupling & $g_e$ & $0.1$ & Perturbative coupling ($g_e^2/4\pi \ll 1$) \\
        Dark matter mass & $m_\chi$ & $10^{-4}$--$10^{2}\,\mathrm{GeV}$ & Sub-MeV to WIMP mass range \\
        Scalar mediator mass & $m_S$ & $10^{-8}$--$10^{2}\,\mathrm{GeV}$ & Light long-range to heavy contact regimes \\
        \bottomrule
    \end{tabular}%
    }
\end{table}

To verify the numerical stability of our piecewise implementation of Eqs.~\eqref{eq:sigma_stable}--\eqref{eq:Phi_series} ($\sigma_{\mathrm{stable}}$) and compare it against the naive double-precision subtraction $I(0) - I(t_{\min})$ in Eq.~\eqref{eq:sigmatot} ($\sigma_{\mathrm{naive}}$) as well as the EFT contact formula~\eqref{eq:sigma_EFT} ($\sigma_{\mathrm{EFT}}$), we compute an independent reference value $\sigma_{\mathrm{ref}}$ using 80-digit arbitrary-precision arithmetic in \textsc{mpmath} (cross-checked against adaptive Gauss--Kronrod quadrature of Eq.~\eqref{eq:dsdt}). Table~\ref{tab:precision} presents the results across nine decades of $m_S$ at $m_\chi = 1\,\mathrm{GeV}$ and $v_{\mathrm{DM}} = 10^{-3}c$ ($q_{\max} = \sqrt{|t_{\min}|} \simeq 1.021\,\mathrm{keV}$). As shown in Table~\ref{tab:precision}, for $m_S \ge 10^{-2}\,\mathrm{GeV}$ ($x \le 10^{-8}$), the naive antiderivative subtraction $\sigma_{\mathrm{naive}}$ loses accuracy rapidly, producing a relative error of $10^{2}$ at $m_S = 1\,\mathrm{GeV}$, $2.6 \times 10^{8}$ at $m_S = 10\,\mathrm{GeV}$, and underflowing to zero at $m_S = 100\,\mathrm{GeV}$. By contrast, $\sigma_{\mathrm{stable}}$ agrees with the 80-digit reference $\sigma_{\mathrm{ref}}$ to within a relative error $|\sigma_{\mathrm{stable}} - \sigma_{\mathrm{ref}}|/\sigma_{\mathrm{ref}} \le 1.2 \times 10^{-15}$ across all rows. Extending this validation across a 420-point grid spanning $m_\chi \in [10^{-4}, 10^{2}]\,\mathrm{GeV}$, $v_{\mathrm{DM}} \in [10^{-4}, 10^{-1}]\,c$, and $m_S \in [10^{-8}, 10^{2}]\,\mathrm{GeV}$ confirms that the relative error of $\sigma_{\mathrm{stable}}$ remains bounded by $\le 1.5 \times 10^{-15}$ throughout the tested domain.

\begin{table}[htbp]
    \centering
    \caption{\label{tab:precision} Numerical validation of the double-precision reformulation $\sigma_{\mathrm{stable}}$ [Eqs.~\eqref{eq:sigma_stable}--\eqref{eq:Phi_series}], naive antiderivative subtraction $\sigma_{\mathrm{naive}}$ [Eq.~\eqref{eq:sigmatot}], and EFT contact approximation $\sigma_{\mathrm{EFT}}$ [Eq.~\eqref{eq:sigma_EFT}] against the 80-digit arbitrary-precision reference $\sigma_{\mathrm{ref}}$ at $m_\chi = 1\,\mathrm{GeV}$, $g_\chi = g_e = 0.1$, and $v_{\mathrm{DM}} = 10^{-3}c$ ($q_{\max} = \sqrt{|t_{\min}|} \simeq 1.021\,\mathrm{keV}$). Here $\varepsilon_{\mathrm{stable}} \equiv |\sigma_{\mathrm{stable}} - \sigma_{\mathrm{ref}}|/\sigma_{\mathrm{ref}}$ and $\varepsilon_{\mathrm{naive}} \equiv |\sigma_{\mathrm{naive}} - \sigma_{\mathrm{ref}}|/\sigma_{\mathrm{ref}}$.}
    \resizebox{\textwidth}{!}{%
    \begin{tabular}{ccccccc}
        \toprule
        $m_S\,[\mathrm{GeV}]$ & $x = |t_{\min}|/m_S^2$ & $\sigma_{\mathrm{stable}}\,[\mathrm{cm^2}]$ & $\varepsilon_{\mathrm{stable}}$ & $\sigma_{\mathrm{naive}}\,[\mathrm{cm^2}]$ & $\varepsilon_{\mathrm{naive}}$ & $\sigma_{\mathrm{EFT}}\,[\mathrm{cm^2}]$ \\
        \midrule
        $10^{-7}$ & $1.043 \times 10^{2}$ & $3.069 \times 10^{-13}$ & $8.2 \times 10^{-16}$ & $3.069 \times 10^{-13}$ & $6.6 \times 10^{-16}$ & $3.233 \times 10^{-11}$ \\
        $10^{-6}$ & $1.043 \times 10^{0}$ & $1.582 \times 10^{-15}$ & $7.5 \times 10^{-16}$ & $1.582 \times 10^{-15}$ & $8.7 \times 10^{-16}$ & $3.233 \times 10^{-15}$ \\
        $10^{-4}$ & $1.043 \times 10^{-4}$ & $3.233 \times 10^{-23}$ & $9.1 \times 10^{-16}$ & $3.233 \times 10^{-23}$ & $1.0 \times 10^{-12}$ & $3.233 \times 10^{-23}$ \\
        $10^{-2}$ & $1.043 \times 10^{-8}$ & $3.233 \times 10^{-31}$ & $9.5 \times 10^{-16}$ & $3.233 \times 10^{-31}$ & $4.6 \times 10^{-6}$  & $3.233 \times 10^{-31}$ \\
        $1.0$     & $1.043 \times 10^{-12}$ & $3.233 \times 10^{-39}$ & $1.2 \times 10^{-15}$ & $3.294 \times 10^{-37}$ & $1.0 \times 10^{2}$   & $3.233 \times 10^{-39}$ \\
        $10.0$    & $1.043 \times 10^{-14}$ & $3.233 \times 10^{-43}$ & $9.9 \times 10^{-16}$ & $8.432 \times 10^{-35}$ & $2.6 \times 10^{8}$   & $3.233 \times 10^{-43}$ \\
        $100.0$   & $1.043 \times 10^{-16}$ & $3.233 \times 10^{-47}$ & $1.1 \times 10^{-15}$ & $0.000$ (underflow)     & $1.0 \times 10^{0}$   & $3.233 \times 10^{-47}$ \\
        \bottomrule
    \end{tabular}%
    }
\end{table}

\begin{figure}[htbp]
    \centering
    \includegraphics[width=0.76\textwidth]{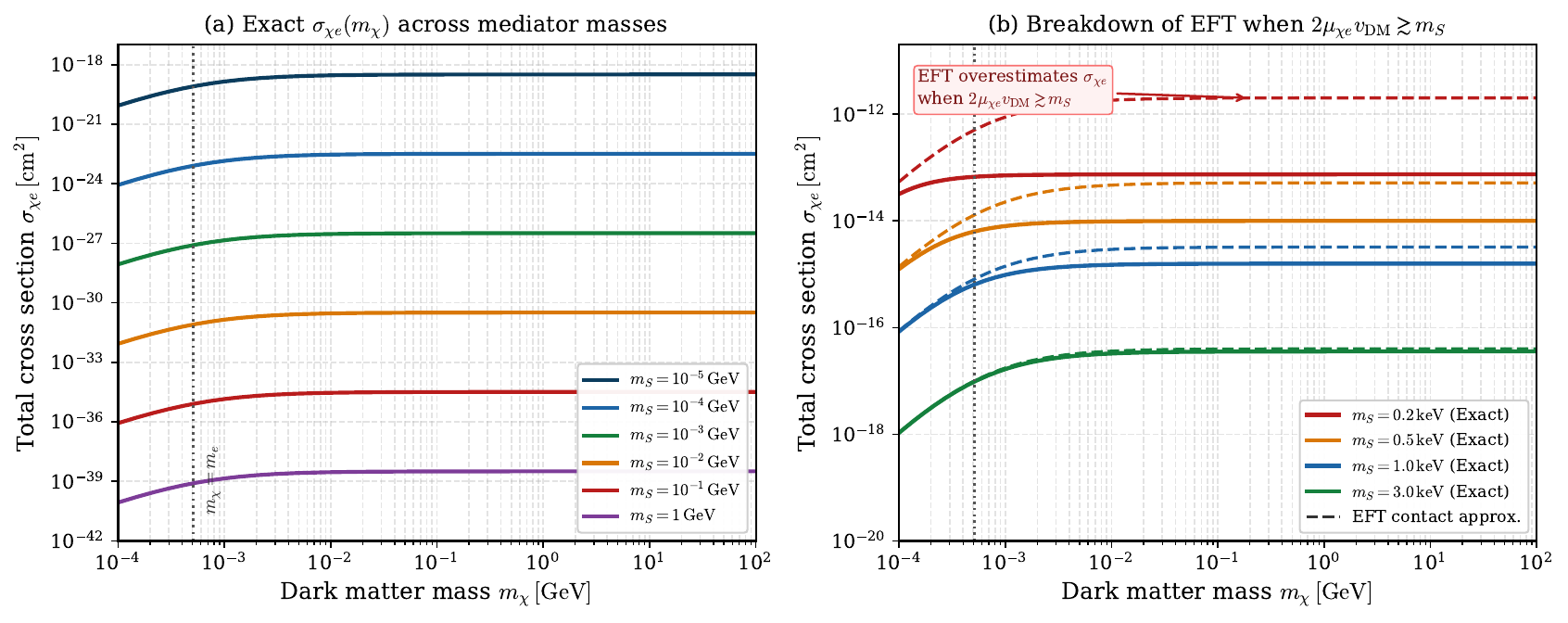}\vspace{-2mm}
    \caption{\label{fig:sigma_mchi} (a) Total elastic cross section $\sigma_{\chi e}$ as a function of dark matter mass $m_\chi$ for scalar mediator masses $m_S \in [10^{-5}, 1]\,\mathrm{GeV}$ with fixed parameters $g_\chi = g_e = 0.1$ and $v_{\mathrm{DM}} = 10^{-3}c$. The vertical dotted line marks $m_\chi = m_e \simeq 0.511\,\mathrm{MeV}$. (b) Comparison between the exact cross section $\sigma_{\mathrm{exact}}$ (solid curves) and the EFT contact approximation $\sigma_{\mathrm{EFT}}$ (dashed curves) for light mediators $m_S \in \{0.2, 0.5, 1.0, 3.0\}\,\mathrm{keV}$ at $g_\chi = g_e = 0.1$ and $v_{\mathrm{DM}} = 10^{-3}c$.}
\end{figure}

Figure~\ref{fig:sigma_mchi}(a) shows the dependence of $\sigma_{\chi e}$ on $m_\chi$ for $10^{-4}\,\mathrm{GeV} \le m_\chi \le 10^{2}\,\mathrm{GeV}$. In the low-mass region $m_\chi \ll m_e$, the reduced mass $\mu_{\chi e} \simeq m_\chi$ grows linearly with $m_\chi$, leading to $\sigma_{\chi e} \propto m_\chi^2$. For $m_\chi \gg m_e$, the reduced mass approaches the constant value $\mu_{\chi e} \to m_e$, and the cross section saturates at a mass-independent plateau. Figure~\ref{fig:sigma_mchi}(b) compares the exact cross section (solid) with the EFT approximation (dashed) for light mediators $m_S \in \{0.2, 0.5, 1.0, 3.0\}\,\mathrm{keV}$. While both descriptions agree when $m_\chi \ll m_e$ ($q_{\max} \simeq 2m_\chi v_{\mathrm{DM}} \ll m_S$), the EFT curves overestimate the cross section by up to two orders of magnitude once $m_\chi \gtrsim m_e$, where $q_{\max} \simeq 2m_e v_{\mathrm{DM}} \simeq 1.02\,\mathrm{keV}$ exceeds $m_S$.

\begin{figure}[htbp]
    \centering
    \includegraphics[width=0.76\textwidth]{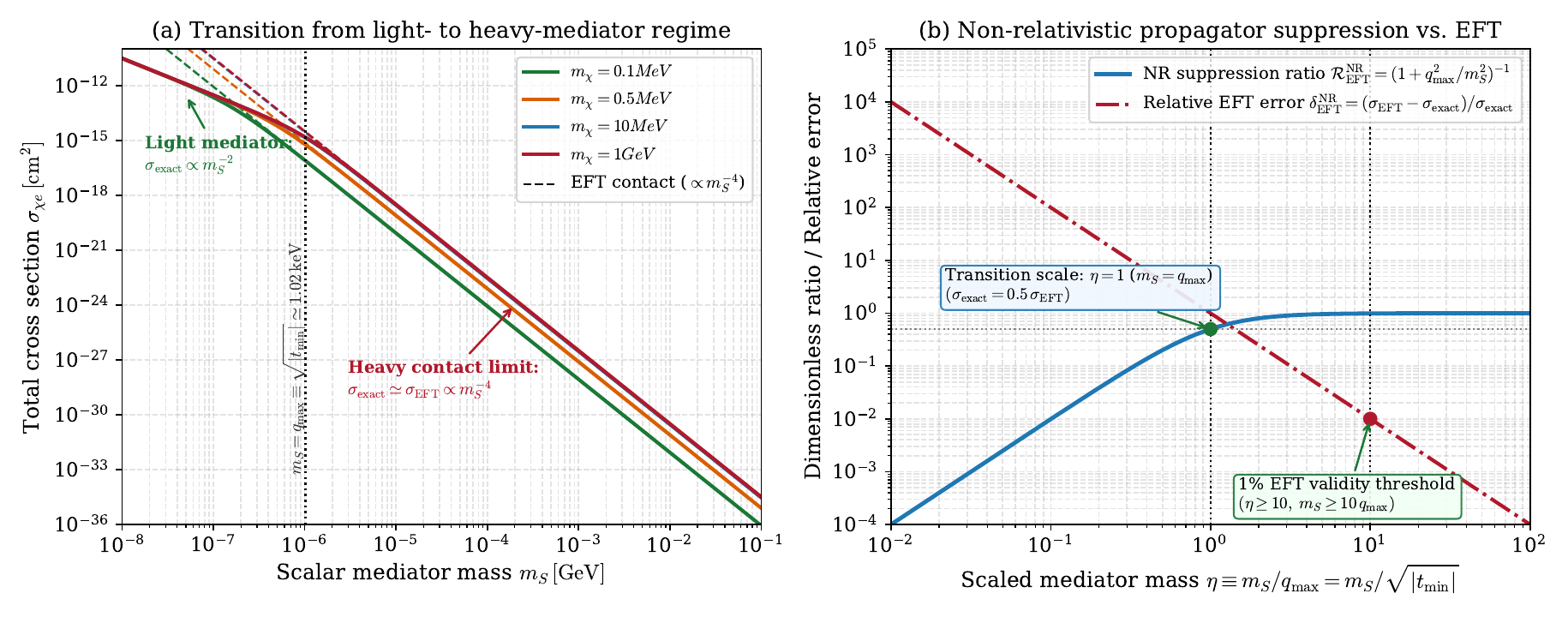}\vspace{-2mm}
    \caption{\label{fig:sigma_mS} (a) Total elastic cross section $\sigma_{\chi e}$ (solid curves) versus scalar mediator mass $m_S \in [10^{-8}, 10^{-1}]\,\mathrm{GeV}$ for $m_\chi \in \{0.1\,\mathrm{MeV}, 0.5\,\mathrm{MeV}, 10\,\mathrm{MeV}, 1\,\mathrm{GeV}\}$ at fixed $g_\chi = g_e = 0.1$ and $v_{\mathrm{DM}} = 10^{-3}c$, compared with the EFT contact limit $\sigma_{\mathrm{EFT}} \propto m_S^{-4}$ (dashed lines). (b) Non-relativistic propagator suppression ratio $\mathcal{R}_{\mathrm{EFT}}^{\mathrm{NR}} = (1 + q_{\max}^2/m_S^2)^{-1}$ (solid blue curve) and relative EFT error $\delta_{\mathrm{EFT}}^{\mathrm{NR}} = (\sigma_{\mathrm{EFT}} - \sigma_{\mathrm{exact}})/\sigma_{\mathrm{exact}}$ (dash-dotted red curve) versus $\eta = m_S/q_{\max} = m_S/\sqrt{|t_{\min}|}$.}
\end{figure}

Figure~\ref{fig:sigma_mS}(a) displays the behavior of $\sigma_{\chi e}$ as a function of $m_S$. For $m_S \gg q_{\max} = \sqrt{|t_{\min}|}$, the curves follow the contact scaling $\sigma \propto \mu_{\chi e}^2 m_S^{-4}$. Below the transition scale $m_S \lesssim \sqrt{|t_{\min}|}$, the propagator softens the slope to $\sigma \propto m_S^{-2} v_{\mathrm{DM}}^{-2}$ in accordance with Eq.~\eqref{eq:sigma_light_NR}, and the curves for different dark matter masses merge toward a common asymptote at fixed $v_{\mathrm{DM}}$. Figure~\ref{fig:sigma_mS}(b) illustrates the non-relativistic suppression ratio $\mathcal{R}_{\mathrm{EFT}}^{\mathrm{NR}}(\eta)$ and relative error $\delta_{\mathrm{EFT}}^{\mathrm{NR}}(\eta)$ from Eq.~\eqref{eq:EFT_ratio}, showing the $50\%$ suppression ($\delta_{\mathrm{EFT}}^{\mathrm{NR}} = 1$) at $\eta = 1$ ($m_S = \sqrt{|t_{\min}|}$) and the $1\%$ convergence threshold at $\eta = 10$ ($m_S = 10\sqrt{|t_{\min}|}$).

\begin{figure}[htbp]
    \centering
    \includegraphics[width=0.48\textwidth]{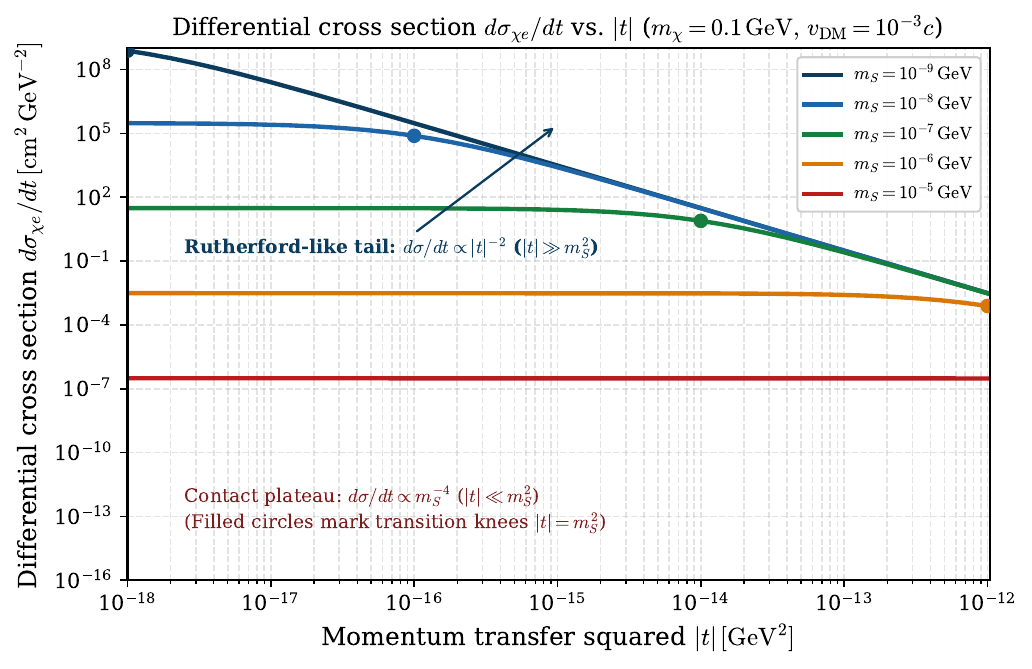}\vspace{-2mm}
    \caption{\label{fig:dsdt} Differential cross section $d\sigma_{\chi e}/dt$ versus squared momentum transfer $|t|$ for mediator masses $m_S \in \{10^{-9}, 10^{-8}, 10^{-7}, 10^{-6}, 10^{-5}\}\,\mathrm{GeV}$ at fixed $m_\chi = 0.1\,\mathrm{GeV}$, $g_\chi = g_e = 0.1$, and $v_{\mathrm{DM}} = 10^{-3}c$ ($|t_{\min}| \simeq 1.04 \times 10^{-12}\,\mathrm{GeV}^2$). Filled circles indicate the transition points $|t| = m_S^2$.}
\end{figure}

Figure~\ref{fig:dsdt} plots the differential cross section $d\sigma_{\chi e}/dt$ as a function of $|t| \in [10^{-18}, 10^{-12}]\,\mathrm{GeV}^2$ for $m_\chi = 0.1\,\mathrm{GeV}$, $g_\chi = g_e = 0.1$, and $v_{\mathrm{DM}} = 10^{-3}c$. For each value of $m_S$, $d\sigma_{\chi e}/dt$ is constant ($\propto m_S^{-4}$) in the low-momentum-transfer regime $|t| \ll m_S^2$, decreases to one-fourth of its plateau value at $|t| = m_S^2$ (filled circles), and approaches the mediator-mass-independent Rutherford-like behavior $d\sigma_{\chi e}/dt \propto |t|^{-2}$ for $|t| \gg m_S^2$.

\section{Direct-Detection Phenomenology and Experimental Bounds}\label{sec:phenomenology}
\subsection{Three-Tier Connection: Microscopic Cross Section, Reference Parameter, and Event Rate}
To connect our analytical results with electron-recoil direct-detection experiments without ambiguity, it is important to distinguish three physical levels of description~\cite{Essig2012,Essig2016,Essig2017}:
\begin{enumerate}\setlength{\itemsep}{1pt}\setlength{\parskip}{0pt}\setlength{\topsep}{2pt}
    \item \emph{Microscopic free-electron differential cross section $d\sigma_{\chi e}/dt$:} Given by Eq.~\eqref{eq:dsdt}, this quantity describes elastic scattering off a free stationary electron in vacuum.
    \item \emph{Reference cross section $\bar{\sigma}_e$ and mediator form factor $F_{\mathrm{med}}(q)$:} Because bound electrons in atoms or semiconductor crystals carry a characteristic orbital momentum $q_0 = \alpha m_e \simeq 3.73\,\mathrm{keV}$ (with $\alpha \simeq 1/137.036$), the standard convention in the sub-GeV direct-detection literature~\cite{Essig2012,Essig2016} factorizes the momentum-transfer dependence of the squared matrix element at $q = \sqrt{-t}$ relative to the fixed reference point $q_0$ ($t_0 = -q_0^2 = -\alpha^2 m_e^2$):
    \begin{equation}
        \bar{\sigma}_e \equiv \frac{\mu_{\chi e}^2}{16\pi\,m_\chi^2 m_e^2}\,\overline{|\mathcal{M}(t = -q_0^2)|^2}, \qquad
        F_{\mathrm{med}}(q) = \frac{q_0^2 + m_S^2}{q^2 + m_S^2},
        \label{eq:sigmabar_def}
    \end{equation}
    where $\overline{|\mathcal{M}(q)|^2} = \overline{|\mathcal{M}(q_0)|^2}\,[(4m_\chi^2 + q^2)/(4m_\chi^2 + q_0^2)][(4m_e^2 + q^2)/(4m_e^2 + q_0^2)]\,|F_{\mathrm{med}}(q)|^2$.
    \item \emph{Macroscopic detector ionization event rate $dR/d\ln E_e$:} The observable differential event rate per unit target mass in a detector is obtained by integrating over the galactic halo velocity distribution $f(\vec{v})$ and the kinematically allowed momentum transfers $q$ weighted by both $|F_{\mathrm{med}}(q)|^2$ and the target-specific atomic or crystal ionization form factor $|f_{\mathrm{ion}}(k', q)|^2$~\cite{Essig2012,Essig2016}:
    \begin{equation}
        \frac{dR}{d\ln E_e} = \frac{\rho_\chi}{m_\chi}\,\frac{\bar{\sigma}_e}{8\mu_{\chi e}^2}\int d^3v\,\frac{f(\vec{v})}{v}\int_{q_{\min}(E_e)}^{q_{\max}} dq\,q\,|F_{\mathrm{med}}(q)|^2\,|f_{\mathrm{ion}}(k', q)|^2,
        \label{eq:rate}
    \end{equation}
    where $\rho_\chi \simeq 0.3\,\mathrm{GeV/cm^3}$ is the local dark matter density and $E_e$ is the electron recoil energy.
\end{enumerate}

\subsection{Analytical Mapping between \texorpdfstring{$d\sigma_{\chi e}/dt$}{dsigma/dt} and \texorpdfstring{$\bar{\sigma}_e$}{sigma_e}}
Combining Eq.~\eqref{eq:sigmabar_def} with our differential cross section~\eqref{eq:dsdt} and the laboratory-frame K\"all\'en identity~\eqref{eq:lambda_exact_lab}, we obtain the dimensionally consistent relation between $\bar{\sigma}_e$ and $d\sigma_{\chi e}/dt$ evaluated at $t_0 = -\alpha^2 m_e^2$:
\begin{equation}
    \begin{aligned}
        \bar{\sigma}_e &= \left.\frac{d\sigma_{\chi e}}{dt}\right|_{t = -\alpha^2 m_e^2} \times \frac{\mu_{\chi e}^2\,\lambda(s, m_\chi^2, m_e^2)}{m_\chi^2 m_e^2} = 4\mu_{\chi e}^2\,\gamma_{\mathrm{DM}}^2 v_{\mathrm{DM}}^2 \left.\frac{d\sigma_{\chi e}}{dt}\right|_{t = -\alpha^2 m_e^2} \\[3pt]
        &= \frac{g_\chi^2 g_e^2\,\mu_{\chi e}^2}{\pi\,(\alpha^2 m_e^2 + m_S^2)^2}\left(1 + \frac{\alpha^2 m_e^2}{4m_\chi^2}\right)\!\left(1 + \frac{\alpha^2}{4}\right).
    \end{aligned}
    \label{eq:sigmabar_exact}
\end{equation}
Here the prefactor $4\mu_{\chi e}^2 \gamma_{\mathrm{DM}}^2 v_{\mathrm{DM}}^2$ has mass dimension $+2$ ($\mathrm{GeV}^2$), compensating the $\mathrm{GeV}^{-2}$ dimension of $d\sigma_{\chi e}/dt$. For sub-GeV and GeV dark matter ($m_\chi \ge 1\,\mathrm{MeV} \gg \alpha m_e \simeq 3.73\,\mathrm{keV}$), the spinor recoil factors inside the parentheses differ from unity by $\alpha^2/4 \simeq 1.33 \times 10^{-5}$ and $\alpha^2 m_e^2/(4m_\chi^2) \le 3.5 \times 10^{-6}$, so Eq.~\eqref{eq:sigmabar_exact} simplifies to better than $2 \times 10^{-5}$ relative accuracy as
\begin{equation}
    \bar{\sigma}_e(m_\chi, m_S, g_\chi g_e) \simeq \frac{g_\chi^2 g_e^2\,\mu_{\chi e}^2}{\pi\,(\alpha^2 m_e^2 + m_S^2)^2}.
    \label{eq:sigmabar_compact}
\end{equation}
Solving Eq.~\eqref{eq:sigmabar_compact} for the effective coupling $\sqrt{g_\chi g_e}$ in terms of $\bar{\sigma}_e$ gives
\begin{equation}
    \sqrt{g_\chi g_e} = \left[\frac{\pi\,\bar{\sigma}_e}{\mu_{\chi e}^2}\right]^{1/4}\sqrt{m_S^2 + \alpha^2 m_e^2}.
    \label{eq:g_bound}
\end{equation}

\subsection{Experimental Inputs, Form-Factor Recasting, and Coupling Bounds}
As evident from Eqs.~\eqref{eq:sigmabar_def} and~\eqref{eq:rate}, an experimental $90\%$ confidence level (C.L.) upper limit $\bar{\sigma}_e^{\mathrm{lim}}(m_\chi)$ extracted from detector count rates depends on the momentum profile of $F_{\mathrm{med}}(q)$ across the integration range $[q_{\min}, q_{\max}]$. Experimental collaborations publish exclusion contours for two standard benchmark limits: (i) the contact form factor $F_{\mathrm{med}}(q) = 1$ (valid when $m_S \gg q \sim \text{few}\times \alpha m_e$), and (ii) the ultra-light mediator form factor $F_{\mathrm{med}}(q) = (\alpha m_e / q)^2$ (valid when $m_S \ll \alpha m_e$). Because typical ionizing transitions in silicon ($E_{\mathrm{gap}} \simeq 1.1\,\mathrm{eV}$) and liquid xenon ($E_{\mathrm{bind}} \simeq 12\,\mathrm{eV}$) are dominated by momentum transfers $q \sim q_{\mathrm{eff}} \equiv \kappa_{\mathrm{det}} q_0$ with $\kappa_{\mathrm{det}} \approx 1.6\text{--}2.8$, a light mediator suppresses the high-$q$ integrand in Eq.~\eqref{eq:rate} by $(q_0/q_{\mathrm{eff}})^4 = \kappa_{\mathrm{det}}^{-4}$, making the published limit $\bar{\sigma}_e^{\mathrm{lim}}[F_{\mathrm{med}} = (\alpha m_e/q)^2]$ weaker than $\bar{\sigma}_e^{\mathrm{lim}}[F_{\mathrm{med}} = 1]$ by a factor of $\kappa_{\mathrm{det}}^4$.

Accordingly, when translating published limits into bounds on $\sqrt{g_\chi g_e}$ across arbitrary $m_S$, we distinguish two procedures:
\begin{enumerate}\setlength{\itemsep}{1pt}\setlength{\parskip}{0pt}\setlength{\topsep}{2pt}
    \item \emph{Fixed-$q_0$ reference projection:} Substituting the heavy-mediator limit $\bar{\sigma}_e^{\mathrm{lim}}[F_{\mathrm{med}}=1]$ directly into Eq.~\eqref{eq:g_bound}. This gives the exact $90\%$ C.L. exclusion bound in the heavy-mediator domain $m_S \gg q_{\mathrm{eff}}$, whereas for $m_S \lesssim \alpha m_e$ it represents a fixed-$q_0$ normalization benchmark.
    \item \emph{Detector-response $q_{\mathrm{eff}}$-recast:} Approximating the detector-averaged form factor in Eq.~\eqref{eq:rate} by $|F_{\mathrm{med}}(q_{\mathrm{eff}})|^2$ where $\kappa_{\mathrm{det}} = q_{\mathrm{eff}}/q_0 = \bigl(\bar{\sigma}_e^{\mathrm{lim}}[F_{\mathrm{med}}\propto q^{-2}] / \bar{\sigma}_e^{\mathrm{lim}}[F_{\mathrm{med}}=1]\bigr)^{1/4}$ matches the published light-mediator and heavy-mediator limits of each experiment, which replaces $\sqrt{m_S^2 + \alpha^2 m_e^2} \to \sqrt{m_S^2 + \kappa_{\mathrm{det}}^2 \alpha^2 m_e^2}$ in Eq.~\eqref{eq:g_bound}.
\end{enumerate}

\begin{figure}[htbp]
    \centering
    \includegraphics[width=0.70\textwidth]{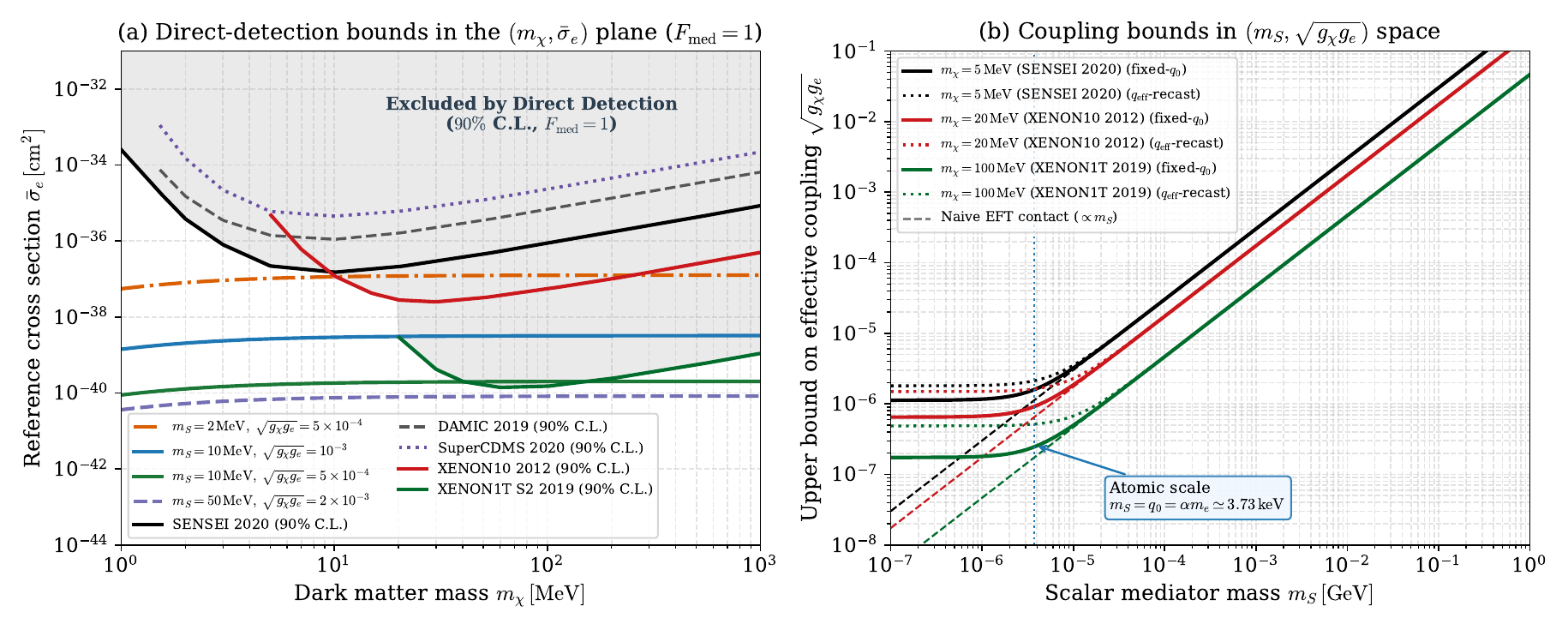}\vspace{-2mm}
    \caption{\label{fig:exclusion} (a) Reference cross section $\bar{\sigma}_e$ versus dark matter mass $m_\chi \in [1, 1000]\,\mathrm{MeV}$ for heavy-mediator benchmarks $m_S \in \{2, 10, 50\}\,\mathrm{MeV}$ ($m_S \gg \alpha m_e$, where $F_{\mathrm{med}}(q) = 1$ holds strictly), compared against published $90\%$ C.L. exclusion limits ($F_{\mathrm{med}}=1$) from SENSEI (2020)~\cite{SENSEI2020}, DAMIC (2019)~\cite{DAMIC2019}, SuperCDMS (2020)~\cite{SuperCDMS2021}, XENON10 (2012)~\cite{XENON10_2012}, and XENON1T S2-only (2019)~\cite{XENON1T_S2_2019}. (b) $90\%$ C.L. upper bounds on $\sqrt{g_\chi g_e}$ versus $m_S \in [10^{-7}, 1]\,\mathrm{GeV}$ for $m_\chi = 5\,\mathrm{MeV}$ (SENSEI), $20\,\mathrm{MeV}$ (XENON10), and $100\,\mathrm{MeV}$ (XENON1T), showing the fixed-$q_0$ projection [Eq.~\eqref{eq:g_bound}, solid curves], the detector-response $q_{\mathrm{eff}}$-recast matching the published $F_{\mathrm{med}}\propto q^{-2}$ bounds (dotted curves), and the naive EFT contact extrapolation (dashed lines).}
\end{figure}

Figure~\ref{fig:exclusion}(a) compares our theoretical predictions for heavy mediators $m_S \in \{2, 10, 50\}\,\mathrm{MeV}$ (where $F_{\mathrm{med}}(q) = 1$ is satisfied to better than $0.1\%$) against the published $90\%$ C.L. exclusion contours ($F_{\mathrm{med}}=1$) digitized from SENSEI (2020)~\cite{SENSEI2020}, DAMIC (2019)~\cite{DAMIC2019}, SuperCDMS (2020)~\cite{SuperCDMS2021}, XENON10 (2012)~\cite{XENON10_2012}, and XENON1T S2-only (2019)~\cite{XENON1T_S2_2019}. Silicon CCD and cryogenic detectors (SENSEI, DAMIC, SuperCDMS) constrain the low-mass region $1\,\mathrm{MeV} \le m_\chi \lesssim 10\,\mathrm{MeV}$ down to $\bar{\sigma}_e \simeq 1.5 \times 10^{-37}\,\mathrm{cm}^2$, while liquid-xenon detectors (XENON10, XENON1T) dominate for $m_\chi \gtrsim 15\,\mathrm{MeV}$, reaching $\bar{\sigma}_e \le 1.5 \times 10^{-40}\,\mathrm{cm}^2$ at $m_\chi = 100\,\mathrm{MeV}$. While the benchmark $(m_S = 10\,\mathrm{MeV}, \sqrt{g_\chi g_e} = 10^{-3})$ is excluded by XENON1T for $m_\chi \gtrsim 22\,\mathrm{MeV}$, $(m_S = 10\,\mathrm{MeV}, \sqrt{g_\chi g_e} = 5 \times 10^{-4})$ and $(m_S = 50\,\mathrm{MeV}, \sqrt{g_\chi g_e} = 2 \times 10^{-3})$ remain allowed across the entire sub-GeV range.

\begin{table}[htbp]
    \centering
    \caption{\label{tab:exp_bounds} Summary of published $90\%$ C.L. upper limits $\bar{\sigma}_e^{\mathrm{lim}}$ for $F_{\mathrm{med}}=1$ and $F_{\mathrm{med}}=(\alpha m_e/q)^2$, together with the resulting upper bounds on $\sqrt{g_\chi g_e}$ across light ($m_S = 0.1\,\mathrm{keV}$, showing both fixed-$q_0$ / $q_{\mathrm{eff}}$-recast values), intermediate ($m_S = 10\,\mathrm{keV}$), and heavy ($m_S = 10\,\mathrm{MeV}$) mediator masses.}
    \resizebox{\textwidth}{!}{%
    \begin{tabular}{llccccc}
        \toprule
        Experiment & $m_\chi$ & $\bar{\sigma}_e^{\mathrm{lim}}(F_{\mathrm{med}}\!=\!1)\,[\mathrm{cm^2}]$ & $\bar{\sigma}_e^{\mathrm{lim}}(F_{\mathrm{med}}\!\propto\!q^{-2})\,[\mathrm{cm^2}]$ & $\sqrt{g_\chi g_e}^{\mathrm{max}}$ ($0.1\,\mathrm{keV}$, $q_0$/recast) & $\sqrt{g_\chi g_e}^{\mathrm{max}}$ ($10\,\mathrm{keV}$) & $\sqrt{g_\chi g_e}^{\mathrm{max}}$ ($10\,\mathrm{MeV}$) \\
        \midrule
        SENSEI (2020)~\cite{SENSEI2020}       & $5\,\mathrm{MeV}$   & $2.2 \times 10^{-37}$ & $1.4 \times 10^{-36}$ & $1.12 \times 10^{-6}$ / $1.80 \times 10^{-6}$ & $3.22 \times 10^{-6}$ & $3.01 \times 10^{-3}$ \\
        DAMIC (2019)~\cite{DAMIC2019}         & $10\,\mathrm{MeV}$  & $1.1 \times 10^{-36}$ & $1.2 \times 10^{-35}$ & $1.64 \times 10^{-6}$ / $2.96 \times 10^{-6}$ & $4.70 \times 10^{-6}$ & $4.40 \times 10^{-3}$ \\
        SuperCDMS (2020)~\cite{SuperCDMS2021} & $10\,\mathrm{MeV}$  & $4.5 \times 10^{-36}$ & $3.8 \times 10^{-35}$ & $2.34 \times 10^{-6}$ / $3.97 \times 10^{-6}$ & $6.68 \times 10^{-6}$ & $6.26 \times 10^{-3}$ \\
        XENON10 (2012)~\cite{XENON10_2012}    & $20\,\mathrm{MeV}$  & $2.8 \times 10^{-38}$ & $7.8 \times 10^{-37}$ & $6.48 \times 10^{-7}$ / $1.49 \times 10^{-6}$ & $1.85 \times 10^{-6}$ & $1.74 \times 10^{-3}$ \\
        XENON1T S2 (2019)~\cite{XENON1T_S2_2019} & $100\,\mathrm{MeV}$ & $1.5 \times 10^{-40}$ & $9.2 \times 10^{-39}$ & $1.74 \times 10^{-7}$ / $4.86 \times 10^{-7}$ & $4.96 \times 10^{-7}$ & $4.65 \times 10^{-4}$ \\
        \bottomrule
    \end{tabular}%
    }
\end{table}

Figure~\ref{fig:exclusion}(b) and Table~\ref{tab:exp_bounds} present the upper bounds on $\sqrt{g_\chi g_e}$ as a function of $m_S$. For $m_S \gg \alpha m_e \simeq 3.73\,\mathrm{keV}$, the fixed-$q_0$ projection, the $q_{\mathrm{eff}}$-recast curve, and the EFT line $\sqrt{g_\chi g_e} \propto m_S$ all coincide. For $m_S \ll 3.73\,\mathrm{keV}$, the naive EFT extrapolation underestimates the allowed coupling by orders of magnitude, whereas the dynamical propagator saturates at a constant plateau ($\sqrt{g_\chi g_e}^{\mathrm{max}} \sim 10^{-7}\text{--}10^{-6}$), with the detector-level $q_{\mathrm{eff}}$-recast relaxing the ultra-light mediator plateau by a factor of $\kappa_{\mathrm{det}} \approx 1.6\text{--}2.8$ relative to the fixed-$q_0$ projection.

\section{Conclusions}\label{sec:conclusions}
In this work, we have investigated tree-level elastic scattering of Dirac dark matter with charged leptons ($\chi + \ell \to \chi + \ell$) mediated by a real scalar boson $S$ in the post-EWSB mass eigenbasis and examined its numerical stability and direct-detection implications:
\begin{enumerate}\setlength{\itemsep}{1pt}\setlength{\parskip}{0pt}\setlength{\topsep}{2pt}
    \item We derived closed-form analytical expressions for the invariant scattering amplitude, spin-averaged squared matrix element $\overline{|\mathcal{M}|^2}$, differential cross sections $d\sigma/dt$ and $d\sigma/d\Omega$, and total cross section $\sigma(s)$, specifying the low-energy effective Lagrangian and its gauge-invariant UV completion via dimension-five Higgs-portal operators.
    \item By combining the factorized laboratory-frame K\"all\'en identity~\eqref{eq:lambda_exact_lab} with the positive-term decomposition~\eqref{eq:sigma_stable} and the convergent power series~\eqref{eq:Phi_series} for the auxiliary recoil integrals $\Phi_1(x)$ and $\Phi_2(x)$, we eliminated the subtractive cancellation affecting naive floating-point evaluations of $I(0) - I(t_{\min})$ when $|t_{\min}| \ll m_S^2$. Comparison against 80-digit arbitrary-precision arithmetic across a 420-point grid spanning $m_\chi \in [10^{-4}, 10^{2}]\,\mathrm{GeV}$, $v_{\mathrm{DM}} \in [10^{-4}, 10^{-1}]\,c$, and $m_S \in [10^{-8}, 10^{2}]\,\mathrm{GeV}$ confirms a relative accuracy of $\le 1.5 \times 10^{-15}$.
    \item We derived both the general and non-relativistic forms of the EFT suppression ratio $\mathcal{R}_{\mathrm{EFT}}$, showing that in the non-relativistic regime ($\mathcal{R}_{\mathrm{EFT}}^{\mathrm{NR}} = (1 + |t_{\min}|/m_S^2)^{-1}$) the contact approximation overestimates the cross section by $100\%$ at the transition scale $m_S = q_{\max} = \sqrt{|t_{\min}|}$ and requires $m_S \ge 10\sqrt{|t_{\min}|}$ for $\le 1\%$ accuracy.
    \item We established a dimensionally consistent relation between $d\sigma_{\chi e}/dt$ and the reference cross section $\bar{\sigma}_e$ at $q_0 = \alpha m_e \simeq 3.73\,\mathrm{keV}$ and evaluated the $90\%$ C.L. coupling bounds from SENSEI, DAMIC, SuperCDMS, XENON10, and XENON1T, comparing fixed-$q_0$ normalization projections against detector-level momentum-transfer recasting across light and heavy mediator masses.
\end{enumerate}

Future work will extend this analytical framework to relativistic scattering of cosmic-ray boosted dark matter ($\gamma_{\mathrm{DM}} \gg 1$), thermal freeze-out relic-density calculations via $s$-channel annihilation $\chi\bar{\chi} \to S \to \ell^+\ell^-$, and one-loop induced photon/Z-boson couplings relevant for complementary nuclear-recoil and collider searches.

\section*{Acknowledgements}
This work was completed building upon scientific materials left by the late Professor Nguyen Suan Han. The author gratefully acknowledges his guidance, mentorship, and lasting scientific influence. Sincere thanks are also extended to the Faculty of Physics and Chemical Engineering, Le Quy Don Technical University for academic support.

\providecommand{\href}[2]{#2}

\end{document}